\documentclass[conference]{IEEEtran}

\ifCLASSINFOpdf
\else
\fi

\usepackage{wrapfig}
\usepackage{threeparttable}
\usepackage{amssymb,amsmath}
\usepackage{graphicx}
\usepackage[ruled,vlined]{algorithm2e}
\usepackage{url}
\usepackage{color}
\usepackage{multicol}
\usepackage{lipsum}
\usepackage{xspace}
\newsavebox{\tablebox}
\usepackage{multirow}
\usepackage{setspace}
\usepackage{paralist}
\let\itemize\compactitem
\let\enditemize\endcompactitem
\let\enumerate\compactenum
\let\endenumerate\endcompactenum

\def\mathbi#1{\textbf{\em #1}}
\newcommand{\next}{\mathbi{X}}
\newcommand{\until}{\mathbi{U}}

 \newcommand{\comment}[1]{}
  \newcommand{\sig}[1]{\textsf{{\small #1}}}

\begin{document}


\title{Formal Consistency Checking over Specifications in Natural Languages}

%
%
\author{

\IEEEauthorblockN{Rongjie Yan}
\IEEEauthorblockA{State Key Laboratory of Computer Science,\\ Institute of Software,\\ Beijing, China
}
\and
\IEEEauthorblockN{Chih-Hong Cheng}
\IEEEauthorblockA{Industrial Software Technologies\\ ABB Corporate Research, \\Ladenburg, Germany
}
\and
\IEEEauthorblockN{Guangquan Zhang and Yesheng Chai}
\IEEEauthorblockA{School of Computer Science \& Technology, \\Soochow University, \\Suzhou, China}
}

\maketitle
\begin{abstract}

Early stages of system development involve outlining desired features such as functionality, availability, or usability. Specifications are derived from these features that concretize vague ideas presented in natural languages.
The challenge for the verification and validation of specifications arises from the syntax and semantic gap between different representations and the need of automatic tools. In this paper, we present a requirement-consistency maintenance framework to produce consistent representations. The first part is the automatic translation from natural languages describing functionalities to formal logic with an abstraction of time.
It extends pure syntactic parsing by adding semantic reasoning and the support of partitioning input and output variables.
The second part is the use of synthesis techniques to examine if the requirements are consistent in terms of realizability. When the process fails, the formulas that cause the inconsistency are reported to locate the problem.

 \end{abstract}

\setcounter{footnote}{0}


\IEEEpeerreviewmaketitle
\vspace{-5pt}
\section{Introduction\label{sec.introduction}}
Early stages of system development involve outlining desired features such as functionality, availability, or usability.
The importance of early verification of high-level requirements can never be underestimated: Precise specifications can avoid overly frequent corrections in late developing phases, and they serve as a reference model or a test-case generator later in system and architecture design. Nevertheless, the challenge for the early verification process arises from the syntax and semantic gap between different representations. We consider a specification to have three views.
Initial specifications are generated from these features that concretize vague ideas presented in natural languages. Precise specifications can be represented under the assist of logic. But for reference models or even test case generators, very often they are represented as programs~\cite{Nguyen2012,Riesco2010}. 
In many cases, one struggles to maintain the following two types of consistency.

\begin{itemize}
\item  The  \emph{semantic} consistency between the intuitive meaning of a textual specification and its logic presentation.
\item The \emph{realizability} consistency between the intuitive meaning of a logic specification and a model or a test case generator - a logic specification should guarantee the existence of an implementation\footnote{E.g., consider the specification ``\emph{the output should always be the same as the input 3 time ticks from now}''. Although it can be rewritten formally using LTL ($ \textbf{G}\,(\sig{output} \leftrightarrow \textbf{XXX}\sig{input})$), it is unrealizable, as any implementation requires to have \emph{clairvoyance} over the future events.}.
\end{itemize}

In this paper, we present a framework that endeavors to bring consistency to different representations of the high-level specification, via a \emph{synthesis-based transformation}. The goal is to generate, from requirements in natural languages, to two other representations automatically, thereby ensuring consistency. We also call it \emph{LTL-oriented}, as the intermediate format makes the use of linear temporal logic (LTL)~\cite{pnueli1977temporal}.

\begin{figure}[t]
\begin{center}
\includegraphics[width=0.8\columnwidth]{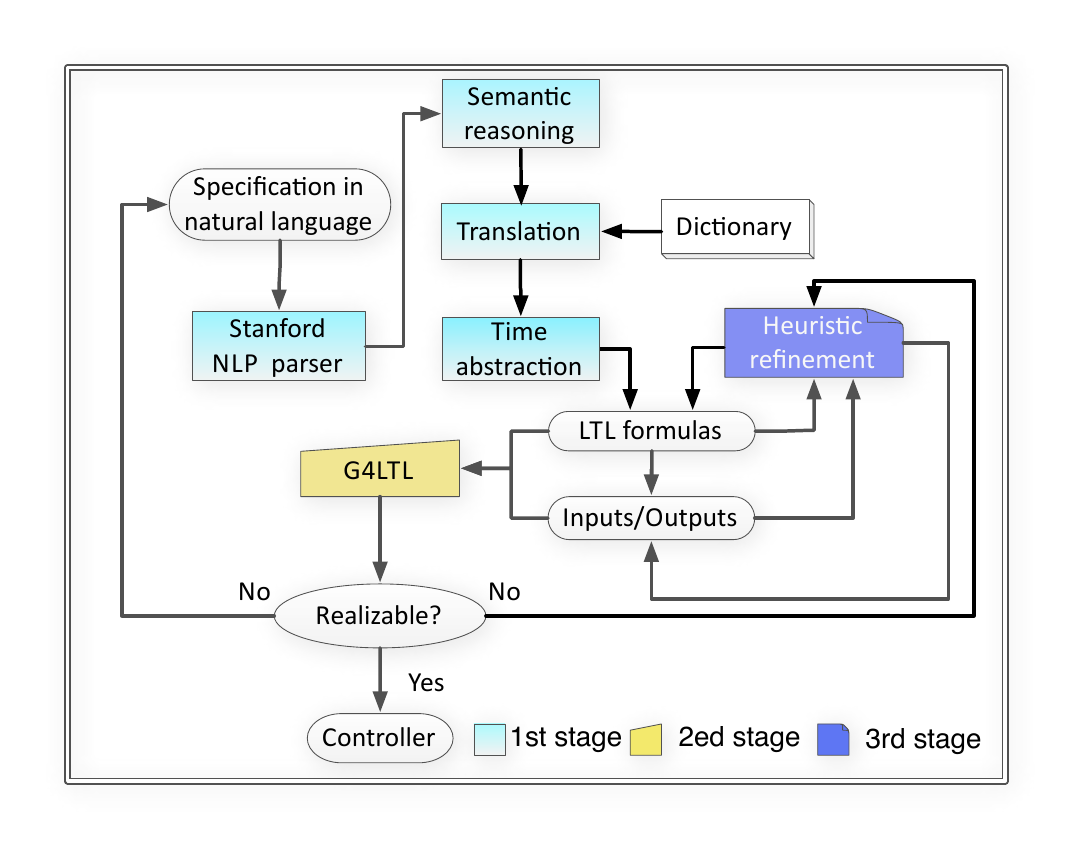}
\end{center}
\vspace{-25pt}
\caption{The overall workflow of SpecCC.\label{fig:workflow}}
\vspace{-15pt}
\end{figure}

The  workflow of our framework (see Figure~\ref{fig:workflow} for overview) involves a loop of three stages. The first part is a translation framework that can automatically turn requirements written in natural language into formulas formalized in LTL. This portion includes a heuristic partition of input and output variables, together with an appropriate abstraction of time. Importantly, apart from pure lexical parsing, we also introduce the use of dictionaries to infer the meaning. This allows the solver to reason status scenarios such as $\sig{on(light)} \equiv \neg \sig{off(light)}$, thereby avoiding the creation of two propositions. The second part is the use of LTL synthesis~\cite{Jobstm06c,acacia12,G4LTLST} to detect the realizability of specifications. The last  part is a semi-automatic procedure that changes the physical interpretation of time and examines if the partition of input-output variables is reasonable. When changes from any of the two items arise, the specification is refined and is re-analyzed by the LTL synthesis engine.

The paper is structured as follows. After a brief summary of related work, we describe the CARA system~\cite{increment3} as a motivating example. We present the translation framework from natural to formal language in Section~\ref{se:concepts}. Then, we summarize the consistency checking between formal language and implementation, together with the following refinement steps in Section~\ref{se:g4ltl}. Section~\ref{se:tool} explains the implementation of the prototype tool SpecCC (\sig{Spec}ification \sig{C}onsistency \sig{C}hecking) and presents the results of applying the tool on a set of non-trivial examples. We conclude the paper in Section~\ref{se:conclusion}.

\vspace{-5pt}
\section{Related Work}\label{se:related}
\vspace{-5pt}

Managing consistencies for requirements of different representations is essential in system modeling and subsequent implementation / refinement steps.

Translating requirements in natural languages to LTL has been investigated in many works, which is always based on a subset of natural language. There are approaches using property patterns and scopes to assist the specification of formal properties specified in LTL as well as other languages~\cite{Dwyer1998,Mondragon2003}.  Machine learning and data mining are also used to classify natural language into temporal requirements~\cite{Nikora2009}. The translator implemented in~\cite{Lukas} is one of the available tools, which translates specifications on hardware systems into LTL for model checking.
In the research of consistency checking of requirements, one of the pioneering work in this domain is~\cite{Heitmeyer1996}, which analyzes requirements expressed in the SCR (Software Cost Reduction) tabular notation.
Li, Liu and He have generated pre and post conditions for system operations of UML requirements model to check requirement consistency based on the semantics~\cite{He2005}. Overall, all above results do not support the consistency checking between formal requirements and implementability. 

In the community of formal methods,  Uchitel, Brunet and Chechik have proposed a method to synthesize a set of safety properties in Fluent Linear Temporal Logic (FLTL) to generate partial behavior models to reduce the effort of model construction~\cite{UchitelBC09}.
One of the closest work from ours is an application inside  robotic domains~\cite{Kress-GazitFP08}, where the technique parses templates to generate LTL formulas and synthesizes a controller. However, the translated specification only supports a strict language subset (GR-1~\cite{piterman2006synthesis}) that disallows the use of \textsf{Until} and strictly limits the use of \textsf{Next} to at most once. Contrarily, we allow the user to still present their requirements in English, while the synthesis engine supports full LTL and allows simple semantic reasoning.
Another similar work~\cite{GhoshELLSS14} translates specifications in stylized natural language into various formalisms from an intermediate representation, and checks the realizability with synthesis analysis with GR-1 synthesis tool. Although the translation distinguishes variables and valuations, time is not considered and valuations have to be converted to the bit-level for synthesis.

\vspace{-5pt}
\section{Example: CARA System}\label{se:exp}
\vspace{-5pt}
We use the CARA (Computer-Aided Resuscitation Algorithm) infusion pump control system \cite{increment3,Ray2004} as a running example.
It is a software system developed to drive a high output infusion pump used for fluid resuscitation of patients suffering from conditions that lead to hypotension, based on the collected data on blood pressure. The main functionality is to monitor and control the operations of the infusion pump, to drive resuscitating fluids into a patient's blood stream.

\noindent\sig{System Description}. We focus on the three main modes (\sig{wait}, \sig{manual}  and \sig{auto-control}) of operation in CARA, and check the consistency of the specification on the three working modes.

The system is in the \sig{wait} mode when the pump is off. It does not monitor the blood pressure or control the pump.
The system enters the \sig{manual} mode when the pump is initially turned on.
In this level, the software only performs monitoring functions.
If the power supply to the pump is lost, the control goes to a backup battery and triggers an alarm.
To leave this mode, either the pump is turned off, or a button for the \sig{auto-control} mode is pressed by the care-giver.
That button is only enabled when the pump is in the normal operative mode.
In the \sig{auto-control} mode, CARA is responsible for monitoring the status lines from the pump and controlling the infusion rate.
When the system is in the \sig{auto-control} mode, it can use three sources of blood pressure data.
Among the sources such as an arterial line, a pulse-wave and a cuff, the arterial line has the highest priority.
That is, if the three sources are both available, the arterial line is used as the input. If the arterial line source is lost, the pulse-wave has a higher priority than the cuff.

The corresponding requirements from~\cite{increment3} are organized as the input of our framework.
The following list enumerates some requirements as illustrating examples.\footnote{A detailed list of requirements is listed in the appendix.}

\begin{small}
\begin{description}
\item [\textsf{Req-08}] If Air Ok signal remains low, auto-control mode is terminated in  3 seconds.
\item [\textsf{Req-17}] When auto-control mode is entered, eventually the cuff will be inflated.
\item [\textsf{Req-28}] If a valid pressure is unavailable in 180 seconds, manual-mode should be triggered.
\item [\textsf{Req-32}] If pulse\_wave or arterial\_line is available, and cuff is selected, corroboration is triggered.
\item [\textsf{Req-42}] When auto-control mode is running, and the arterial\_line or pulse\_wave  or cuff is lost, an alarm should sound in 60 seconds.
\item [\textsf{Req-44}] If pulse\_wave and arterial\_line are  unavailable, and cuff is selected, and blood\_pressure is not valid, next manual\_mode is started.
\end{description}
\end{small}

\vspace{-10pt}
\section{Maintaining Consistencies between Natural Language and Formal Language}\label{se:concepts}
\vspace{-5pt}
Maintaining consistencies between natural language and formal logic is obtained by automatic translation from natural language to LTL.
In the following subsections, we review the definition of LTL, propose a restricted English grammar for syntactic parsing, and give the underlying algorithm for an extended reasoning on the semantic level and the techniques for abstracting time.
Afterwards, we provide heuristics  on partitioning input and output variables to apply the synthesis techniques in the second level.

\vspace{-10pt}
\subsection{LTL Syntax and Semantics}
\vspace{-5pt}
Linear temporal logic~\cite{pnueli1977temporal} is a modal temporal logic whose modalities refer to time.
It can express properties of paths in a computation tree.
The formulas are built up from a set of atomic propositions (AP), logical operations and temporal modal logics. The set of supported LTL formulas are constructed from atomic propositions $p\in AP$ according to the following grammar.
$$\phi::=p\; |\; \neg  \phi \; |\; \phi \vee \phi \; |\;\next \phi \;|\; \lozenge \phi \;|\; \square \phi \;| \;\phi \until \phi \; ,$$
where $\next$ is the \textsf{Next} time operator, $\lozenge$ is the \textsf{Eventually} operator, $\square$ is the \textsf{Always} operator, and $\until$ is the \textsf{Until} operator. Given negation ($\neg$) and disjunction ($\vee$), we can define conjunction ($\wedge$), implication ($\rightarrow$) and equivalence ($\leftrightarrow$).
The semantics of an LTL formula is defined on an infinite sequence of truth assignments to the atomic propositions. For example,
a temporal logic formula $p$ is satisfiable if there is a sequence of states such that $p$ is true from the start of the sequence.

\vspace{-10pt}
\subsection{Grammar of the Structured English}

Natural languages allow a rich diversity of sentence structures and easily cause semantic ambiguities.
Hence, we do not provide a translator for the full set of natural languages.
Instead, we select a more structured subset, whose constituting grammar can encompass the needs of functional requirements in most cases.
In this subset, we only support present, future and passive tenses with correct syntax according to the English grammar. To be concise, we only present positive form of the grammar. The negative form can be inferred accordingly.
We first present the grammar of the structured English.

\[
\scriptsize
\begin{array}{lll}
{sentence} & ::= & (subclause, )^*. (clauses).(, subclause)^*\\
{subclause} & ::= & {(subordinator)}. {(clauses)} \\
{clauses} & ::= & (clause). [, (conjunction).(clause)] \\
{clause} & ::= &[modifier]. {(subject)} . {(predicate)}. \\ && [constraint]\\
subject & ::= &substantives\\
substantives &::= & (substantive). [(conjunction).\\&&(substantives)] \\
substantive & ::= & noun \\
predicates & ::= &[modality].predicate\\
predicate & ::= & verb\;|\; (be).participle \;|\; (be).(complement)\\
participle & ::= & (verb).(\text{ed}) \;|\; (verb).(\text{ing})\\
complement & ::= & adjective \;|\; adverb\\
modality & ::= & \text{shall} \;| \text{should}\; |\text{will} \; | \text{would}\;| \text{can}\;| \text{could} \; | \text{must}\\
 {subordinator} & ::= &\text{if}\; |\; \text{after}\;|\;\text{once} \;|\;\text{when} \;|\;\text{whenever}\; |\; \text{while}\;\\&& |\;  \text{before}\; |\; \text{until}\; |\; \text{next}\\
 modifier & ::= &\text{globally} \;|\; \text{always} \;|\; \text{sometimes} \;|\; \text{eventually}\\
conjunction & ::= & \text{and} \;|\; \text{or}\\
constraint&::= & \text{in } t \\
\end{array}
\]
where $t$ is a time constant, * (star) means the presence of zero or more subcomponent, . (dot) means
the composition of different components, and [ ] means the optional components. For \emph{noun}, \emph{verb}, \emph{participle}, \emph{adjective} and \emph{adverb}, 
we do not decompose them any more. 
The proposed grammar allows extracting temporal operators according to the elements in \emph{subordinator} and \emph{modifier}.

According to the definition of the grammar, a sentence consists of at least one clause.
A clause consists of at least one main phrase for the core meaning of the sentence, and optionally conditional or time clauses that extend the meaning.
In addition to the skeleton shown in the above grammar, a main phrase may also contain filter constructions such as \sig{the} and \sig{a} that will be ignored in the translation. 
For example, for Requirement \sig{Req-17}:  ``\sig{ When auto-control mode is entered, eventually the cuff will be inflated.}'', the main clause is ``\sig{eventually the cuff will be inflated}'', and ``\sig{When auto-control mode is entered}'' is a time clause. In this requirement, ``\sig{eventually}'' is a modifier for the temporal operator \sig{Eventually}  and ``\sig{the}'' is a filter construction in the main phrase ``\sig{eventually the cuff}''.

\vspace{-5pt}
\subsection{Lexical and Syntactic Parsing}

A specification here is a set of sentences (requirements) in the structured English.
For every sentence, first it is parsed by a natural language parser to extract all grammatical ingredients. Then according to the dependency relation extracted by the parser, the translator decomposes the sentence into clauses recursively to isolate the independent temporal units.
After the decomposition, we construct a syntax tree for the ingredients of the sentence, to extract atomic propositions, and
to infer the temporal relationship of individual temporal elements according to the subordinators and modifiers.
Usually an atomic proposition comes from a subject and its predicate extracted by the dependency relation from the parser, i.e.,  in the form of \emph{predicate\_subject}, to combine a variable and its valuation. For multiple subjects connected with conjunctions, they are decomposed to generate different atomic propositions and then connected by the corresponding logic operators.

\begin{figure}[t]
\begin{center}
\includegraphics[width=0.9\columnwidth]{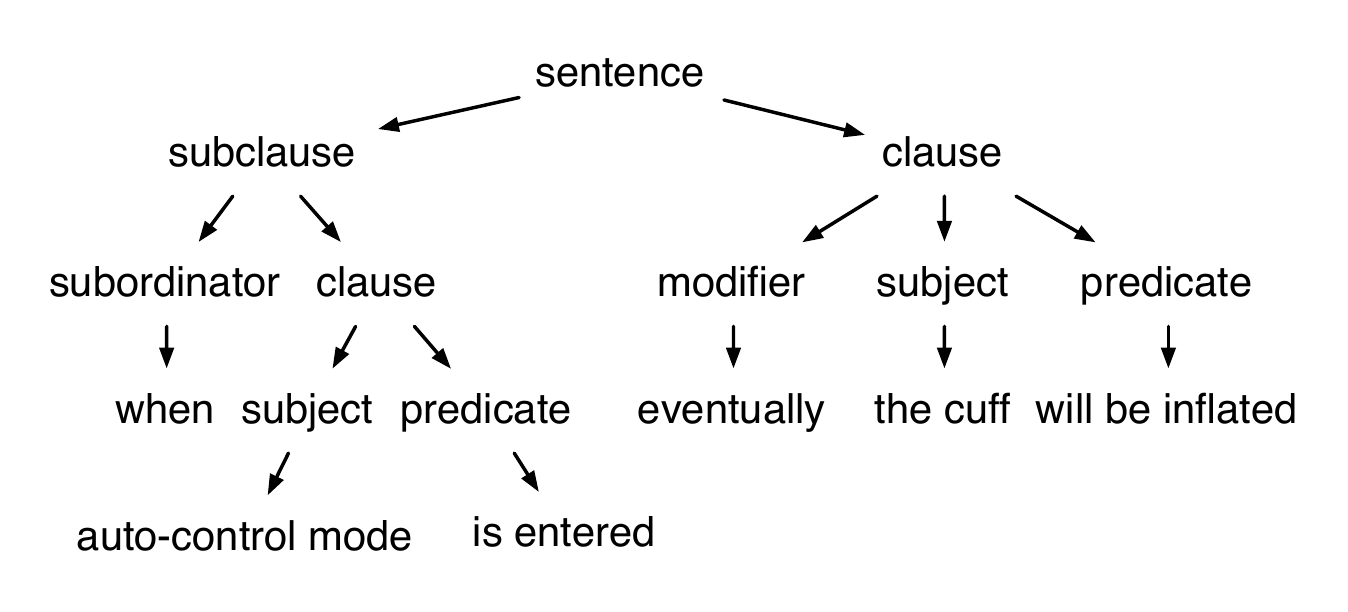}
\end{center}
\vspace{-20pt}
\caption{The syntax tree of Requirement \sig{Req-17}\label{fig:syntaxtree}}
\vspace{-15pt}
\end{figure}

Reconsider the example used in the last subsection. Requirement \sig{Req-17} consists of two clauses.  The corresponding syntax tree is illustrated in Figure \ref{fig:syntaxtree}. By removing tense information in the predicates, two atomic propositions, \sig{enter\_auto-control\_mode} and \sig{inflate\_cuff} are extracted from the two clauses.
Note that only one word will be taken as the subject in a clause in the translation process. To keep every proposition meaningful, we need to add ``\_'' to contact relative words together for a complete subject in a requirement.

According to the existing study on the scopes and patterns, and their templates on LTL formula generation~\cite{Salamah2005},
we have selected some frequently used patterns (Universality and Existence), and the corresponding templates in our translator,
to transform the elements in the syntax tree to the corresponding formulas.
For example, according to the syntax tree for \sig{Req-17} and its keywords,
modifier ``eventually'' is translated into unary temporal operator ``$\lozenge$'', which will be joined with atomic proposition \sig{inflate\_cuff}. Subordinator ``when'' is translated into the logic operator ``$\rightarrow$'', to connect the two formulas from the two clauses. Accordingly,
the temporal formula for \sig{Req-17} is $\square (\sig{enter\_auto-control\_mode} \rightarrow \lozenge \sig{inflate\_cuff})$.


\vspace{-10pt}
\subsection{Moving from Lexical Parsing towards Semantic Reasoning}\label{se:semantic}
To preserve semantic consistency and obtain succinct temporal formulas, additional to pure lexical parsing,
we also adopt semantic reasoning over specifications in natural language.
The semantic reasoning discussed here is to extract the relation between adjectives and adverbs (we call them \emph{antonym candidates} in the rest of the paper) respectively according to their meaning in a specification,
instead of semantic roles labelled by a parser.
More precisely, after extracting the antonym candidates used in a specification, we group them in pairs of
semantically contrasting words by looking up an antonym dictionary specified by users.
With these semantically related words, we can reduce the number of atomic propositions used in the generated formulas, and avoid adding the assumptions on the mutual exclusive propositions.

The most direct way is to look for the antonyms online for every given word,
which is easy to be implemented but time consuming.
We propose a two-step antonym extraction process to compute the pairs of antonyms in a specification.
Intuitively, pairs of antonyms always come from the sentences with same subjects.
Therefore, we first organize the antonym candidates related to the same subjects according to their dependency relation $\langle subject, dependent\rangle$ extracted by the natural language parser, where every word in the dependent set is initialized with color \sig{green}.
If the number of words in the dependent set for a subject is larger than one, we use an antonym dictionary 
to check whether semantically contrasting words exist in the same set. Otherwise, we continue to deal with other groups. The reason being that we cannot use the derived antonyms for the corresponding proposition reduction.
The details on the reasoning process are depicted in Algorithm \ref{alg:semantic}, where \sig{subject} is to group elements depending on the same subjects, and \sig{wordset} is to store the set of antonym candidates  with the extracted antonyms. To mark the status of dependent words in a subject during semantic reasoning, we use two colors, where blue for having found the words in the dependent set of the subject with contrasting meaning by consulting the dictionary,
and green stands for the non-existence of antonyms in the set. These colors are indicators for our proposition reduction in the formula transformation process. That is, a word marked with green will be directly converted into atomic propositions with the corresponding subject. Otherwise, for pairs of antonyms related to the same subject, we first decide the positive and negative properties. Then we replace  the words in negative meaning with the negative form of their antonyms.

\begin{algorithm}[t]
\begin{scriptsize}
\DontPrintSemicolon
\SetKwRepeat{doUntil}{do}{until}
\SetKwInOut{Input}{input}\SetKwInOut{Output}{output}
\Input{Specification $S$ in natural language}
\Output{pairs of antonyms in the specification}
\Begin{
\nl $\textsf{wordset}= \emptyset$\;
    \tcp{extract subject dependent words, and store antonym candidates and the dependency relations in \textsf{wordset} and \textsf{subject}, respectively}
\nl  $\textsf{subject}  = extract(S,\textsf{wordset})$\;
    \For{$\forall s\in \textsf{subject}$}{
\nl    \If{$|s.dep|>1$}{
        \For{$\forall w \in  s.dep$}{
        	\If{$w.color==green$}{
\nl             \If{$\textsf{wordset}(w).antonym == \emptyset$}{
\nl                 $\textsf{wordset}(w).antonym = online(w)$\;}
\nl             $\textsf{antonmy} =s.dep \cap \sig{wordset}(w).antonym$\;
\nl             \If{$\textsf{antonmy}\neq \emptyset$}{
		 $w.color = blue$\;
%
               		 \For{$w'\in \sig{antonym}$}{
\nl                     		$w'.color = blue$\;
\nl                    		 $\textsf{wordset}(w').antonym = \textsf{wordset}(w').antonym \cup \{w\}$\;}
        }
	
    }
    }
    }
}
    return (\textsf{subject, wordset})
    }
\caption{Semantic reasoning\label{alg:semantic}}
\end{scriptsize}
\end{algorithm}

In Algorithm \ref{alg:semantic}, we assume that every antonym candidate can find its antonyms from an antonym dictionary. We first initialize \sig{wordset} (Line 1) and extract the dependency relation from the specification (Line 2). In the \emph{extract} function, the extracted antonym candidates are stored in \sig{wordset}, where initially the sets of their antonyms are empty. The items  in the dependent sets of subjects are initialized with green. Then for every extracted word $w$ in subject $s$,
if it is not analyzed before (Line 4), we
look for its antonyms from the antonym dictionary and add the result to the antonym set of $w$ in \sig{wordset} (Line 5). If it has been processed, we check whether the intersection between its antonyms and the set of words in $s.dep$ is empty (Line 6). If the interaction is not empty, we say that there exists a pair of antonyms in $s.dep$, and mark the corresponding words in $s.dep$ with blue (Lines 7 and 8). Finally, we will complete the information for the antonyms of $w$ (Line 9).

Consider Requirements 32 and 44 presented in Section \ref{se:exp} as an example.
The subject \sig{pulse\_wave} has dependency relation with \sig{available}, and \sig{unavailable}. Without semantic reasoning, we will create two propositions as \sig{available\_pulse\_wave}, and \sig{unavailable\_pulse\_wave}.
In our semantic reasoning process,
we need to look up the antonyms for \sig{available} and \sig{unavailable} in the antonym dictionary. 
As the intersection between \sig{available} and the set of antonyms for \sig{unavailable} is not empty, we say that the two words constitute a pair of antonyms. Therefore, we can replace the proposition \sig{unavailable\_pulse\_wave} by $\neg $ \sig{available\_pulse\_wave}.
Note that we cannot reason antonyms for verbs. The reason being that different actions capture different semantics and we cannot simply use their negation to replace the corresponding antonyms.

\vspace{-8pt}
\subsection{Time Counting and Abstraction}\label{se:time}

Requirements in natural language may contain timing constraints. In LTL, one can use \textsf{Next} operator to emulate discrete time.
That is, decide unit time (e.g., 1~second), and define the elapse of unit time via a \textsf{Next} operator.
For example, for Requirement  \textsf{Req-08}:``\sig{If Air Ok signal remains low, auto-control mode is terminated in  3 seconds}'', its LTL formula is $\square (\neg \sig{Air\_ok\_signal }  \rightarrow XXX \sig{terminate\_auto-control\_mode})$, where one $\next$ is for one second.

Explicit enumerating time units brings a clearer mapping for requirement analysis. However, the larger the number of time units in a requirement, the longer of the translated formula with a higher complexity.
To alleviate this problem,
we present a rewriting technique that can abstract time with the consideration of all the requirements inside the specification.
Intuitively, we can reduce the numbers of \textsf{Next} operators by dividing them with the \emph{greatest common divisor (GCD)} for the lengths of successive \textsf{Next}  operators in the requirements. The method is sound, i.e., a specification with the greatest common divisor reduction is realizable, if and only if the original specification is realizable.

For example, in Requirements \sig{Req-08},  \sig{Req-28} and  \sig{Req-42} in Section \ref{se:exp}, the lengths for successive \sig{Next} operators are 3, 180 and 60, respectively. The greatest common divisor among the three numbers is 3. Therefore, after reduction, the lengths of the successive \sig{Next} operators for the three requirements are 1, 60 and 20, respectively. So the corresponding formula for \sig{Req-08} becomes $
\square (\neg \sig{Air\_ok\_signal }  \rightarrow X \sig{terminate\_auto-control\_mode})$.
From this example, one can observe that the reduction is quite conservative and still produces formula with huge amount of \sig{Next}, thereby hindering the later synthesis process.

The complete algorithm overcomes the limitation via the introduction of \emph{arrival errors}, borrowed from the concept of jitter\footnote{Jitter is the undesired deviation from true periodicity of an assumed periodic signal in electronics and telecommunications.}. Intuitively, an arrival error specifies, for an action, whether it is allowed for arriving later or earlier. This allows us to search for a better re-encoding of consecutive  \sig{Next} operators. With arrival errors, one can formulate the problem into an optimization problem (nonlinear with degree at most~2).

Formally, let $\Theta=\{\theta_0,\ldots, \theta_n\}$ be a set of numbers of successive \textsf{Next} operators in the transformed formulas from a specification where $\theta_i\neq \theta_j$ for any $i\neq j$. Let $d$ be the devisor, $\theta'_i$ be the length of successive \textsf{Next} operators after the abstraction from $\theta_i$, and $\Delta_i$ be the error introduced in the abstraction. With these notations, we obtain the following constraint system, for all $i \in \{0, \ldots, n\}$,
\begin{equation}\label{eq:time}
\theta_i = \theta'_i\times d + \Delta_i, -d < \Delta_i < d, d \text{ and }\theta'_i\in \mathbb{Z}^+, \text{and }  \Delta_i\in \mathbb{Z},
\end{equation}
where $\mathbb{Z}$ is the set of integers, and $\mathbb{Z}^+$ is the set of positive integers including zero.
Intuitively, $\theta_i = \theta'_i\times d + \Delta_i$ means that the algorithm rewrites $\theta_i$ consecutive \textsf{Next} operators into $\theta'_i$  \textsf{Next} operators, via
a factor of $d$. This generates an error of $\Delta_i$.
If a $\Delta_i$ is positive, it means that the corresponding event happens earlier in the rewritten specification. Otherwise, the event arrives later than expected. In the above GCD case,  $\Delta_i$ should always be~$0$.

Based on the above constraint system, we derive a two-objective optimization problem: Reduce the use of \textsf{Next} (i.e., reduce $\theta'_i$), and minimize the controlled error (i.e., reduce $|\Delta_i|$, where $|\Delta_i |$ stands for the absolute value of $\Delta_i$).
\begin{equation}\label{eq:theta}
\text{minimize }\sum^n_{i=0} \theta'_i, ~~~\text{minimize }\sum^n_{i=0} |\Delta_i|
\end{equation}

As the first objective is more important, we let the user specify an allowed upper-bound $B$ for $\sum^n_{i=0} |\Delta_i|$, and restrict the domain for every $\Delta_i$ such that either $\Delta_i \in [0, d]$ or $\Delta_i \in [-d, 0]$ (i.e., arrival error can be either early or late, but not both).  Then the second objective can be rewritten as a single
integer linear constraint (as one can remove the absolute sign). This reduces the two-objective optimization problem into a single-objective optimization problem, which can be efficiently solved by modern SMT solvers (e.g., Yices 2~\cite{DBLP:conf/cav/Dutertre14}) via bit-blasting.

Back to the example and consider again Requirements \textsf{Req-08},  \textsf{Req-28} and  \textsf{Req-42} in Section \ref{se:exp}. The set for the lengths of \textsf{Next} operators in these requirements is $\{3, 180,60\}$, that is, $\theta_0=3, \theta_1=180$ and $\theta_2=60$.
If we specify $\Delta_i\geq 0$ for $0\leq i\leq 2$, and set $B$ to $5$, we could obtain an optimal result with $d=60$, and
\[\begin{array}{llllll}
\theta'_0 = 0, & \theta'_1 = 3, &\theta'_2 = 1, &\Delta_0 = 3, &\Delta_1 = 0, &\Delta_2 = 0.  \\
\end{array}
\]

\vspace{-10pt}
\subsection{Input and Output Variable Partition}

To check the consistencies of a specification,
the translator also needs to distinguish the propositions for input and output variables in an LTL formula.
Very commonly, we find that every line of a specification discusses how the system should response, when encountering a certain scenario. This can be understood as having an implication between an occurring event and the corresponding response. Sometimes, the response might need to endure until another environment event appears. Generally speaking, for left-hand parts in an implication, or for right-hand parts of the \sig{Until} operator, we assume that the constituting variables are input variables. If a proposition in positive form appears in the both sides of such operators, it is assumed as an output.

We use this idea to analyze every requirement and maintain, for each requirement, two disjoint sets for input and output variables. Then when all requirements are analyzed, a unification is performed by checking consistencies among variable sets. Whenever there exists a conflict (i.e., a variable is considered to be an input variable in one requirement but an output variable in another requirement), the translator regards it as an output.
If no input variable exists after the partition, we select one randomly from the set of output variables as the input.
After the partition, the translator also asks the user for the assistance. For example, for the generated formula from \sig{Req-32}: $\square$((available\_pulse\_wave $||$ available\_arterial\_line) \&\& select\_cuff $\rightarrow$ trigger\_corroboration),
we can conclude that  \sig{available\_pulse\_wave}, \sig{available\_arterial\_line} and \sig{select\_cuff } are the input variables, and the output variable is \sig{trigger\_corroboration}.


\vspace{-10pt}
\section{Maintaining Consistencies between Formal Language and Implementability}\label{se:g4ltl}

\subsection{Synthesis}

In real world applications, there are always a large amount of
requirements describing functionality of the desired system. The
inconsistency between various requirements may arise between
different sequences of events (actions) defined by the semantics of
the corresponding LTL formulas. As the synthesis technique is
capable of generating a controller over a set of LTL formulas, the
inconsistency between different LTL formulas will prevent the
synthesis tool from creating a controller. Therefore, we can borrow
the idea of LTL-based synthesis techniques to check the
inconsistency among the requirements represented in LTL formulas.

\vspace{-5pt}
\subsection{Heuristic refinement}

We use \sig{G4LTL}~\cite{G4LTLST} as our underlying LTL synthesis engine,
which automatically checks consistency between LTL formulas transformed from the
requirements in a specification, and reports the inconsistency
between neighbored requirements.
However, it cannot always provide reasons causing the inconsistency,
especially when the pair of requirements causing inconsistencies are not neighbored.
We suggest the following strategy to isolate the problem.
\begin{itemize}
\item Locate the pair of inconsistent requirements.
The process starts from a subset of consistent formulas.
We can add more formulas continuously to the subset to check which one is not consistent with the subset.
Once we have located the problem,
we could filter out other formulas that do not contain any propositions of the located formulas.
\item Adjust the existing input and output variable partition.
As the input and output partition method is heuristic,
we may have to adjust the partition if \sig{G4LTL} reports the inconsistency.
The propositions belonging to the intermediated variables in the located formulas are targets to be adjusted.
\item Modify the requirements.
When we cannot guarantee the consistency by adjusting the partition on input and output variables,
the specification is not consistent indeed.
Therefore, we analyze the semantics of the formulas to extract the reason of inconsistency,
and modify the requirements to ensure the consistency of the specification.
\end{itemize}

\vspace{-10pt}
\section{Implementation and Evaluation}\label{se:tool}
\vspace{-5pt}

We have implemented a research prototype  for the above mentioned framework.
For the language translation part, the implementation extends the Stanford NLP parser~\cite{marneffe06} to analyze the requirements in the structured English to extract its ingredients for LTL generation.
In the consistency checking part, we integrate the synthesis library \textsf{G4LTL}~\cite{G4LTLST}.

We have selected three case studies coming from different application areas, to maintain the consistencies in the two levels, and check the specification scalability that we can handle in the second level.

\begin{description}
\item [CARA] We first have considered Requirements 1, 7, 8, 13, 16, 17, 20, 28, 32, 34, 42, 44, 48, 49 and 54 among seventy requirements that are related to the switches between working modes and the corresponding responsibilities  for the CARA system as the input of our framework.  We also adopted the detailed specifications for the three components (Pump Monitor, Blood Pressure Monitor (BPM), Polling Algorithms (PA)) in the CARA system for evaluation.

\item [TELEPROMISE] We have also considered the functional specification for the TELEPROMISE project demonstrator as a case study\footnote{Available at: www.eng.utoledo.edu/eecs}, which covers five generic applications: Shopping application, Article processing application, On-line reservation application, Information application and Local bulletin board application.

\item [Rescue robot] We have modified the rescue robot scenario used in \cite{Kress-GazitFP08} as the third case study. The responsibility of the robots in this scenario is to look for the injured people and take them to a medic who is in some room. Different numbers of rooms and robots have been considered here, with the constraint that two robots cannot be in the same room at the same time.

\end{description}

Table \ref{tab:result} lists the results for the corresponding component specifications and the scale of every specification (number of lines, and number of I/Os), as well as the time consumption (in seconds) for realizability checking of requirements in LTL formulas.  The performance of \sig{G4LTL} are sensitive  to the number of subformulas, the number of input and output variables, and the length of a formula. Therefore, the time consumption for realizability checking of various specifications are quite different. For the consistency maintenance between natural language and formal language, the time consumption is linear to the number of requirements, which is ignored here.

In the CARA case study, the specifications for the transformation between three working modes and the three components are consistent.
During consistency checking for the TELECOMPRISE case study, \sig{G4LTL} failed to generate controllers for the last two specifications. The failure was caused by the classification of input and output variables.
After locating the problem and modifying the input/ output variable partition, the specifications are consistent.

\vspace{-10pt}
\section{Conclusion}\label{se:conclusion}\vspace{-5pt}

\begin{table}[t]
\centering
\caption{Experimental results \label{tab:result}}
\begin{lrbox}{\tablebox}
\begin{tabular}{l|   l lllll}\hline
Name & No. & Specification & formulas & in & out &time(s) \\\hline
\multirow{14}{*}{CARA} &
0 & Working mode and switching & 30 & 22 & 28   & 34\\\cline{2-7}
&1 & Pump Monitor & 20 & 9&14  &2 \\ \cline{2-7}
&2.1.1 & BPM: cuff detector & 14 & 13 & 12 & 1\\
&2.1.2 & BPM: AL detector & 15 & 11 & 14 & 2\\
&2.1.3 & BPM: pulse wave detector & 14 & 9 & 12 & 1\\
&2.2.1 & BPM: initial auto control & 16 & 14 & 15 & 1\\
&2.2.2 & BPM: first corroboration & 19 & 11 & 16 & 29\\
&2.2.3 & BPM: valid ctrl blood pressure & 13 & 11 & 10 & 2\\
&2.2.4 & BPM: cuff source handler & 11 & 9 & 10 & 2\\
&2.2.5 & BPM: arterial line blood pressure & 16 & 9 & 13 &1 \\
&2.2.6 & BPM: arterial line corroboration & 12 & 8 & 13 & 1\\
&2.2.7 & BPM: pulse wave handler & 20 & 10 & 21 & 23 \\
\cline{2-7}&3.1 & (PA) Model ctrl algorithm & 9 & 15  & 11 & 3\\
&3.2 & (PA) Polling algorithm & 56 & 12& 20  & 11\\\hline
\hline
\multirow{5}{*}{TELE} &1 & Shopping &29 & 11&24  &8 \\
&2& Article processing &  17 & 3 &13  & 1\\
&3 & On-line reservation & 6 & 3 & 4 & 1\\
&4 & Information &15  &  8&  14&1 \\
&5 &Local bulletin board & 17 & 7 &16 & 1\\\hline\hline
\multirow{3}{*}{Robot} & 1 & A  robot with 4 rooms &9 &2 &5 &1 \\
& 2 & A robot with 9 rooms & 14&2  & 10 &1\\
&3 &Two robots with 5 rooms &25 & 2 &11&7\\\hline

\end{tabular}
\end{lrbox}
\scalebox{0.75}{\usebox{\tablebox}}
\end{table}

We have proposed a framework for ensuring consistencies between different representations of specifications. The framework maintains semantic consistencies between oral and formal specifications, while ensuring the implementability using synthesis.
With the framework, we propose a time extraction mechanism, input-output partition heuristics, and an extension from pure syntactic parsing of sentences to reason on the semantics level. The approach is evaluated under a rich set of examples with positive results.

For future work, we will increase the maturity level of our implementation by capturing the semantics in natural language to better understand the meaning of sentences thus avoiding manual matching between different words, and by using learning techniques for more generic natural languages.

\vspace{-10pt}
\bibliographystyle{abbrv}


\newpage
\appendix

\section{Appendix}
\subsection{A full list of checked requirements for the CARA system, together with the automatically generated LTL formulas}


\begin{small}
\begin{description}
 \item [\textsf{Req-01}]
The CARA will be operational whenever the LSTAT is powered on.
\item [\textsf{Req-07}] 
If an occlusion is detected, and auto\_control\_mode is running, auto\_control\_mode will be terminated.
\item [\textsf{Req-08}] 
If Air\_Ok\_signal remains low, auto\_control\_mode is terminated in 3 seconds.
\item [\textsf{Req-13.1}] 
If arterial\_line and pulse\_wave are corroborated, and cuff is available, next arterial\_line is selected.
\item [\textsf{Req-13.2}]
If pulse\_wave is corroborated, and cuff is available, and arterial\_line is not corroborated, next pulse\_wave is selected.
\item [\textsf{R-13.3}] If arterial\_line is not corroborated, and pulse\_wave is not corroborated, and cuff is available, then cuff is selected.
\item [\textsf{Req-16}]
If a pump is plugged in, and an infusate is ready, and the occlusion\_line is clear, auto\_control\_mode can be started.
\item [\textsf{Req-17.1}]
When auto\_control\_mode is running, eventually the cuff will be inflated.
\item [\textsf{Req-17.2}] If start\_auto\_control\_button is pressed, and cuff is not available, an alarm is issued and override\_selection is provided.
\item [\textsf{Req-17.3}] If alarm\_reset\_button is pressed, the alarm is disabled.
\item [\textsf{Req-17.4}] If override\_selection is provided, if override\_yes is pressed, and arterial\_line is not corroborated, next arterial\_line is selected.
\item [\textsf{Req-17.5}] If override\_selection is provided, if override\_yes is pressed, and arterial\_line is corroborated, and pulse\_wave is not corroborated, next pulse\_wave is selected.
\item [\textsf{Req-17.6}] If override\_selection is provided, if override\_no is pressed, next manual\_mode is started.
\item [\textsf{Req-17.7}] If cuff and arterial\_line and pulse\_wave are not available, next manual\_mode is started.
\item [\textsf{Req-20}]
If manual\_mode is running and start\_auto\_control\_button is pressed, next corroboration is triggered.
\item [\textsf{Req-28}]
If  a valid blood\_pressure is unavailable in 180 seconds, manual\_mode should be triggered.
\item [\textsf{Req-32.1}]
If pulse\_wave or arterial\_line is available, and cuff is selected, corroboration is triggered.
\item [\textsf{Req-32.2}] If pulse\_wave is selected, and arterial\_line is available, corroboration is triggered.
\item [\textsf{Req-34}]
When auto\_control\_mode is running, terminate\_auto\_control\_button should be available.
\item [\textsf{Req-42}]
When auto\_control\_mode is running, and the arterial\_line, or pulse\_wave  or cuff is lost, an alarm should sound in 60 seconds.
\item [\textsf{Req-44}]
If pulse\_wave and arterial\_line are  unavailable, and cuff is selected, and blood\_pressure is not valid, next manual\_mode is started.
\item [\textsf{Req-48.1}]
Whenever termiante\_auto\_control\_button is selected, a confirmation\_button is available.
\item [\textsf{Req-48.2}] If a confirmation\_button is available, and confirmation\_yes is pressed, manual\_mode is started.
\item [\textsf{Req-48.3}] If a confirmation\_button is available, and confirmation\_no is pressed, auto\_control\_mode is running.
\item [\textsf{Req-48.4}] If a confirmation\_button is available, and confirmation\_yes is pressed, next confirmation\_yes is disabled.
\item [\textsf{Req-48.5}] If a confirmation\_button is available, and confirmation\_no is pressed, next confirmation\_no is disabled.
\item [\textsf{Req-48.6}] If a confirmation\_button is available, and terminate\_auto\_control\_button is pressed, next terminate\_auto\_control\_button is disabled.
\item [\textsf{Req-49}]
When a start\_auto\_control\_button is enabled, the start\_auto\_control\_button is enabled until it is pressed.
\item [\textsf{Req-54}]
If auto\_control\_mode is running, and impedance\_reading is unavailable, next auto\_control\_model is terminated.
\end{description}
\end{small}

We have applied semantic reasoning and the  abstraction technique on time constants.
As there are no explicit positive and negative meaning for a pair of antonyms, the selection for the positive form is randomly.
When there is only one pair of adjective or adverb  antonyms for a subject, we  abbreviate the propositions by just using the subject and its negative form. For example, available\_pulse\_wave is rewritten as pulse\_wave, and $\neg$ available\_pulse\_wave is rewritten as $\neg$ pulse\_wave. However, other propositions with the same subject will not be modified.

The translated LTL formulas are as follows:
\begin{small}
\begin{description}
 \item [\textsf{Req-01}]
$\square$(power\_lstat $\rightarrow$ ($\lozenge$(operational\_cara)))
 \item [\textsf{Req-07}]
$\square$(detect\_occlusion \&\& run\_auto\_control\_mode $\rightarrow$ ($\lozenge$(terminate\_auto\_control)))
 \item [\textsf{Req-08}]
$\square$($\neg$air\_ok\_signal $\rightarrow$ terminate\_auto\_control\_mode)
 \item [\textsf{Req-13.1}] 
 $\square$(corroborate\_arterial\_line \&\& corroborate\_pulse\_wave \&\& cuff $\rightarrow$ select\_arterial\_line)
 \item [\textsf{Req-13.2}] $\square$(corroborate\_pulse\_wave \&\& cuff \&\& $\neg$corroborate\_arterial\_line $\rightarrow$ select\_pulse\_wave)
 \item [\textsf{Req-13.3}]
 $\square$($\neg$corroborate\_arterial\_line \&\& $\neg$corroborate\_pulse\_\\wave \&\& cuff $\rightarrow$ select\_cuff)
 \item [\textsf{Req-16}]
$\square$(plug\_pump \&\& ready\_infusate \&\& clear\_occlusion\_line $\rightarrow$ start\_auto\_control\_mode)
 \item [\textsf{Req-17.1}] 
$\square$(run\_auto\_control\_mode $\rightarrow$ ($\lozenge$(inflate\_cuff)))
 \item [\textsf{Req-17.2}] $\square$(press\_start\_auto\_control\_button \&\& $\neg$cuff $\rightarrow$ issue\_alarm \&\& provide\_override\_selection)
 \item [\textsf{Req-17.3}] $\square$(press\_alarm\_reset\_button $\rightarrow$ $\neg$alarm)
 \item [\textsf{Req-17.4}] $\square$(provide\_override\_selection $\rightarrow$ ($\square$(press\_override\_yes \&\& $\neg$corroborate\_arterial\_line $\rightarrow$ select\_arterial\_line)))
 \item [\textsf{Req-17.5}] $\square$(provide\_override\_selection $\rightarrow$ ($\square$(press\_override\_yes \&\& corroborate\_arterial\_line \&\& $\neg$corroborate\_pulse\_wave $\rightarrow$ select\_pulse\_wave)))
 \item [\textsf{Req-17.6}] $\square$(provide\_override\_selection $\rightarrow$ ($\square$(press\_override\_no $\rightarrow$ start\_manual\_mode)))
 \item [\textsf{Req-17.7}] $\square$($\neg$cuff\&\&$\neg$arterial\_line\&\&$\neg$pulse\_wave $\rightarrow$ start\_\\manual\_mode)
 \item [\textsf{Req-20}]
$\square$((run\_manual\_mode \&\& press\_start\_auto\_control\_button) $\rightarrow$ trigger\_corroboration)
 \item [\textsf{Req-28}]
$\square$(XXX($\neg$blood\_pressure)$\rightarrow$ trigger\_manual\_mode)
 \item [\textsf{Req-32.1}]
$\square$((pulse\_wave $||$ arterial\_line) \&\& select\_cuff $\rightarrow$ trigger\_\\corroboration)
 \item [\textsf{Req-32.2}] $\square$(select\_pulse\_wave \&\& arterial\_line $\rightarrow$ trigger\_\\corroboration)
 \item [\textsf{Req-34}]
$\square$(run\_auto\_control\_mode $\rightarrow$ terminate\_auto\_control\_\\button)
 \item [\textsf{Req-42}]
$\square$(run\_mode \&\& ($\neg$arterial\_line $||$ $\neg$pulse\_wave $||$ $\neg$cuff) $\rightarrow$ (X(sound\_alarm)))
 \item [\textsf{Req-44}]
$\square$($\neg$pulse\_wave\&\&$\neg$arterial\_line \&\& select\_cuff \&\& $\neg$blood\_pressure $\rightarrow$ start\_manual\_mode)
 \item [\textsf{Req-48.1}]
$\square$(select\_termiante\_auto\_control\_button $\rightarrow$ confirmation\_\\button)
 \item [\textsf{Req-48.2}]$\square$(confirmation\_button \&\& press\_confirmation\_yes $\rightarrow$ start\_manual\_mode)
 \item [\textsf{Req-48.3}]$\square$(confirmation\_button \&\& press\_confirmation\_no $\rightarrow$ run\_auto\_control\_mode)
 \item [\textsf{Req-48.4}]$\square$(confirmation\_button \&\& press\_confirmation\_yes $\rightarrow$ $\neg$confirmation\_yes)
 \item [\textsf{Req-48.5}]$\square$(confirmation\_button \&\& press\_confirmation\_no $\rightarrow$ $\neg$confirmation\_no)
 \item [\textsf{Req-48.6}]$\square$(confirmation\_button \&\& press\_terminating\_auto\_\\control\_button $\rightarrow$ $\neg$terminating\_auto\_control\_button)
 \item [\textsf{Req-49}]
$\square$(start\_auto\_control\_button $\rightarrow$ ($\square$($\neg$press\_start\_auto\_\\control\_button $\rightarrow$ (start\_auto\_control\_button W press\_start\_\\auto\_control\_button))))
 \item [\textsf{Req-54}]
$\square$(run\_auto\_control\_mode \&\& $\neg$impedance\_reading $\rightarrow$ terminate\_auto\_control\_model)
\end{description}
\end{small}

\end{document}